\documentstyle[bo99,epsfig]{article}

\title{THE RELATIVISTIC PRECESSION MODEL FOR QPOS
IN LOW MASS X-RAY BINARIES}
\author{L. Stella}
\affil{Osservatorio Astronomico di Roma, Via Frascati 33,
I-00040 Monteporzio Catone (Roma), Italy,
e-mail stella@coma.mporzio.astro.it; affiliated to I.C.R.A.}

\begin{document}

\maketitle

\begin{abstract}

The relativistic precession model for quasi periodic oscillations, QPOs,
in low mass X-ray binaries is reviewed. The behaviour of three
simultaneous types of QPOs is well matched in terms of the fundamental
frequencies for geodesic motion in the strong field of
the accreting compact object for reasonable star masses and spin
frequencies. The model
ascribes the higher frequency kHz QPOs, the lower frequency kHz
QPOs and the horizontal branch oscillations to the Keplerian, periastron
precession and nodal precession frequencies of matter inhomogeneities
orbiting close to the inner edge of the accretion disk.
The remarkable correlation between the centroid frequency of QPOs
in both neutron star and black hole candidate low mass X-ray binaries
is very well fit by the model.
QPOs from low mass X-ray binaries might provide an unprecedented
laboratory
to test general relativity in the strong field regime.
\end{abstract}

\section{Introduction}
Old accreting neutron stars, NSs, in low mass X-ray binaries, LMXRBs,
display
a complex variety of quasi-periodic oscillation, QPO, modes in their
X-ray
flux. The {\it low frequency} QPOs ($\sim 1-100$~Hz)
that were discovered and studied from high luminosity Z-sources in the
eighties are further classified into horizontal, normal and flaring
branch oscillations (HBOs, NBOs and FBOs, respectively), depending on the
simultaneous position occupied by a source in the X-ray colour-colour
diagram (for a review see van der Klis 1995).
The kHz QPOs ($\sim 0.2$ to $\sim 1.3$~kHz) that were
revealed and investigated with RXTE in a number of NS LMXRBs
(see van der Klis 1998, 1999, 2000 and references therein)
involve timescales similar to the dynamical timescales close to the NS.
A common phenomenon is the presence of a pair of
kHz QPOs (centroid frequencies of $\nu_1$ and $\nu_2$)
which drift in frequency while mantaining their
frequency difference $\Delta\nu \equiv \nu_2 - \nu_1 \approx 250-360$~Hz
roughly constant. Detailed studies showed that in four sources  
$\Delta\nu$ decreases significantly (by up to 
$\sim 100$~Hz) as $\nu_2$ increases; these are 
Sco~X-1 (van der Klis et al. 1997), 4U1608-52 (Mendez et al. 1998a,b),
4U1735-44 (Ford et al. 1998) and 4U1728-34 (Mendez \& van der Klis
1999).
Owing to poor statistics, a similar variation of $\Delta\nu$ in other
sources would have remained undetected (Psaltis et al. 1998).

kHz QPOs show remarkably similar properties
across NS LMXRBs of the Z and Atoll groups, the luminosity
of which differs by a factor of $\sim 10$ on average.
During type I bursts from six Atoll
sources, a nearly coherent signal at a frequency of
$\nu_{burst} \sim 290 - 580$~Hz
has also been detected (for a review see 
Strohmayer 2000). In a few cases $\nu_{burst}$ is
consistent, to within the errors, with the frequency separation of the
kHz QPO pair $\Delta\nu$ or twice its value $2\Delta\nu$.
Yet there are currently two sources (4U1636-53, Mendez et al. 1999,
and 4U1728-34,  Mendez \& van der Klis 1999) for which $\nu_{burst}$ is
significantly different from $\Delta\nu$ and its harmonics.

The presence of HBOs has been firmly established in both Atoll and Z-sources.
Their frequency, $\nu_{HBO}$ ($\sim 15$ to $\sim 60$~Hz)
shows an approximately quadratic
dependence ($\sim \nu_2^2$) on the higher kHz QPO frequency that is observed
simultaneously in a number of sources. 
The frequency changes of the kHz QPOs and HBOs
are positively correlated with the instantaneous accretion rate.
Some evidence has also been found for an equivalent of the NBOs and FBOs
of Z-sources
(Wijnands et al. 1999; Psaltis, Belloni \& van der Klis 1999).

A remarkable correlation between the centroid frequencies of QPOs
(or peaked noise components) from LMXRBs
has been recently discovered (Psaltis, Belloni \& van der Klis 1999). 
This correlation extends over nearly 3 decades in frequency and 
encompasses both NS and black hole candidate, BHC, systems.
The frequencies of these QPOs, despite their quasi-periodic nature, 
provide the most accurately measured observables of LMXRBs. 
A primary goal of any QPO model is therefore to explain the frequency 
range and dependence of the different QPO types of 
these sources. 
The basic features of the relativistic precession model, RPM,
are reviewed here
(Stella \& Vietri 1998a, 1999; Stella, Vietri \& Morsink 1999).
In the RPM the QPO signals arise from the fundamental frequencies
of the motion of matter in the vicinity of the NS.
The corresponding orbits are supposed to be slightly eccentric and
tilted.
As in other models, the higher frequency kHz QPOs at
$\nu_2$ are produced by the $\phi$-motion ({\it i.e.} the Keplerian
motion) of
inhomogeneities orbiting the inner disk boundary, while the
lower frequency QPO signal at $\nu_1$ originates from their periastron
precession, which is primarily determined by strong-field effects.
The HBOs are due to the nodal precession in the orbits of the same
inhomogeneities, an effect which is dominated by frame-dragging
around fast-rotating collapsed stars.
The RPM can be applied to BHCs as well.

\section{Periastron Precession and kHz QPOs}

We consider here only infinitesimally eccentric and tilted orbits,
under the assumption that the motion of matter in the innermost disk
regions is dictated by the star's
gravity alone. In the case of a circular geodesic in the equatorial
plane
($\theta=\pi/2$) of a Kerr black hole
of mass $M$ and specific angular momentum $a$,
the coordinate frequency measured by a static observer at infinity is
\begin{equation}
\nu_\phi=\pm M^{1/2}r^{-3/2} [2\pi (1\pm a M^{1/2}r^{-3/2})]^{-1} \ \
\end{equation}
(we use units such that $G=c=1$).
The upper sign refers to prograde orbits.
If we slightly perturb a circular orbit
in the $r$ and $\theta$ directions, the coordinate frequencies of
the small amplitude oscillations within the plane (the epicyclic
frequency $\nu_r$) and in the perpendicular direction (the vertical
frequency $\nu_\theta$) are given by (see Stella \& Vietri 1999
and references therein)
\begin{equation}
\nu_r^2=\nu_\phi^2 (1-6M r^{-1} \pm 8aM^{1/2} r^{-3/2}- 3a^2 r^{-2}) \ \
,
\end{equation}
\begin{equation}
\nu_\theta^2=\nu_\phi^2 (1\mp 4 aM^{1/2}r^{-3/2}+3a^2r^{-2}) \ \ .
\end{equation}
In the Schwarzschild limit ($a=0$) $\nu_\theta$ coincides with
$\nu_\phi$, such that the nodal precession frequency
$\nu_{nod} \equiv \nu_\phi - \nu_\theta$ is identically  zero.
$\nu_r$, instead, is always lower than the other two frequencies,
reaching
a maximum for $r=8M$ and going to zero at $r_{ms}=6M$.
This qualitative behaviour of $\nu_r$ is preserved in the
Kerr field ($a \neq 0$). Therefore the periastron precession frequency
$\nu_{per}\equiv \nu_\phi-\nu_r$ is dominated by a ``Schwarzschild" term
over a wide range of parameters (Stella \& Vietri 1999).

In the RPM the higher and lower frequency kHz QPOs are identified
with $\nu_2=\nu_\phi$ and $\nu_1=\nu_{per}$, respectively.
Therefore $\Delta\nu \equiv \nu_2-\nu_1 = \nu_{\phi}
- (\nu_{\phi} - \nu_{r}) = \nu_{r}$.
For $a=0$, Eqs. 1-2 give
\begin{equation}
\nu_r=\nu_\phi(1-6M/r)^{1/2} = \nu_\phi[1-6(2\pi\nu_\phi M)^{2/3}]^{1/2}
\ \ .
\end{equation}
The curves in Fig.~1A show $\nu_{r}$ vs. $\nu_{\phi}$ for $a=0$ and
selected values of $M$, the
only free parameter in Eq.~4.
The measured $\Delta\nu$ vs. $\nu_2$  for eleven NS LMXRBs is also
plotted.
It is apparent that for NS masses in the 2~M$_{\odot}$ range,
the simple model outlined above is in qualitative agreement with the
measured
values, including the  decrease of $\Delta\nu$ for increasing $\nu_2$
seen in Sco~X-1, 4U1608-52, 4U1735-44 and 4U1728-34.
The model above is only an
approximation: first, the spacetime around a fast rotating NS
is different from a Kerr spacetime (due to the star's oblateness induced
by rotation);
second, the orbits might possess a finite
(though small !) eccentricity.
\begin{figure}
\centerline{\psfig{file=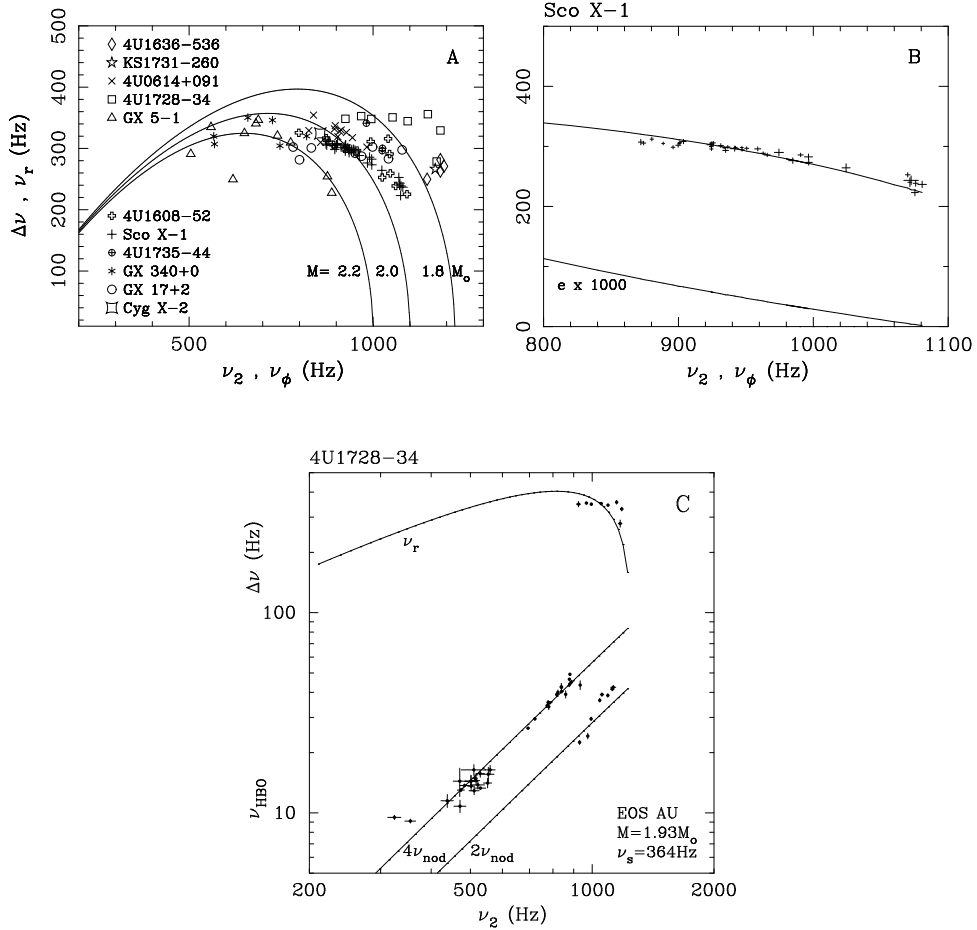, width=12cm, angle=-90}}
\caption[]{ (A) kHz QPO frequency difference $\Delta\nu$ versus higher
QPO frequency $\nu_2$ for eleven LMXRBs.
Error bars are not plotted for the sake of clarity. The curves give
the $r$- and $\phi$-frequencies of matter in nearly circular orbit
around a
non-rotating neutron star, of mass 2.2, 2.0 and 1.8 M$_{\odot}$.
(B) $\Delta\nu$ versus $\nu_2$ in Sco X-1.
The best fit model corresponds to the $r$- and $\phi$-frequencies
of matter orbiting a non-rotating 1.90~M$_{\odot}$ neutron star
at a periastron distance of 6.25~$M$ (17.5 km). The line marked
with $e$ gives the orbital eccentricity ($\times 1000$)
as a function of the $\phi$-frequency.
(C) kHz QPO frequency difference $\Delta\nu$
and (double-branched) HBO frequency versus higher QPO frequency $\nu_2$
in 4U1728-34 (Strohmayer et al. 1996; Ford \& van der Klis 1998;
Mendez \& van der Klis 1999).
The solid lines give the $r$-frequency and the 2nd and 4th
harmonics of the nodal precession frequency $\nu_{nod}$
as a function of the
$\phi$-frequency for infinitesimally eccentric and tilted orbits
in the spacetime of a 1.93~M$_{\odot}$ neutron star spinning at
364~Hz (EOS AU; Wiringa et al. 1988).}
\end{figure}
Analytical formulae to partly correct for these effects were
derived by Stella \& Vietri (1999); 
Fig.~1B shows the fit to the observed $\Delta \nu$ vs. $\nu_2$
relationship in Sco~X-1 that was obtained through them.
The orbital eccentricity, in particular,
was varied in order to obtain different frequencies, while keeping the
periastron distance $r_p = a (1-e)$ fixed. The
best model for a non-rotating NS is shown in
Fig.~1B.
The model reproduces
fairly accurately the data with a minimum number of free parameters,
the NS mass ($M\sim1.9$~M$\odot$) and periastron
distance ($r_p \simeq 6.2$~$M$). The latter value is
close to the marginally stable orbit radius.
When the NS spin is allowed a finite value (say $\nu_s \sim 300-600$~Hz),
fits of very similar quality are obtained, the parameters of which
differ only slightly from those given above. In essence, the effects
induced by the NS rotation on $\nu_r$ are small, though non-negligible.

The behaviour of the curves in Fig.~1A,B , 
and therefore the ability of the model to match the observations,
reflects the properties of the
strong field Schwarzschild metric, since lower order expansions
fail to reproduce the observed frequencies (see Stella \& Vietri 1999).

Within the RPM, the maximum value of $\nu_r = \Delta\nu$ depends mainly
on
the mass of the compact object. The NS masses deduced from the simple
modelling
in Fig.~1A,B are in the $\sim 1.8-2.0$~M$_\odot$ range, in agreement
with the
only relatively accurate mass measurement from optical
spectro-photometry
in any of these systems (Cyg~X-2; $M = 1.78\pm 0.23$~M$_\odot$ ; Orosz \&
Kuulkers 1999).
In general within the RPM, $\Delta\nu$ should not be obviously related to 
the NS spin frequency $\nu_s$.
Therefore, it seems natural to identify $\nu_s$ with $\nu_{burst}$, {\it
i.e.}
the  stable frequency seen during type I X-ray bursts
(for a review see Strohmayer 2000).
The distribution of NS spins inferred in this way
is fairly wide ($\sim 290 - 580$~Hz) and compares well with that of
millisecond radio pulsars, MSPs, in agreement with evolutionary scenarios 
in which LMXRBs are the progenitors of MSPs. The $401$~Hz spin of
SAX~J1808.4-3658, the only bursting LMXRB displaying
coherent pulsations in its persistent emission
(Wijnands \& van der Klis 1998; Chakrabarty \& Morgan 1998), is also in the 
range of spin frequencies deduced from $\nu_{burst}$.
None of the LMXRBs of the (high luminosity) Z-class has yet displayed
burst oscillations; therefore their spin period is still to be measured.
Cyg~X-2 and GX~17+2, the only  type I X-ray bursters in the group, might
provide this important piece of information.

\section{Nodal Precession and HBOs}

If the orbits giving rise to the signals at $\nu_\phi$ and
$\nu_{per}$ are slightly tilted  relative to the equatorial
plane, nodal precession will take place around the spin axis.
In the RPM the HBO frequency is
related to the nodal precession frequency. From Eqs.~1 and 3
this can be written in the slow rotation limit ($a/M \ll 1$)
\begin{equation}
\nu_{nod} \simeq
4\pi a \nu_{\phi}^2 \simeq
6.2\times 10^{-5} (a/M) m \nu_{\phi}^2 \ {\rm Hz}
\simeq 4.4\times10^{-8}\ I_{45}m^{-1} \nu_{\phi}^2 \nu_{s} \ {\rm Hz} \
,
\end{equation}
where $M=m$~M$_\odot$.
This is the well known Lense-Thirring nodal precession formula.
The latter equality refers to a rotating NS, where $aM=2\pi \nu_s I$
with
$I=10^{45}I_{45}$~g~cm$^2$ its moment of inertia.

If $\nu_\phi$ and $\nu_s$ are measured, the only parameter
in Eq.~5 that is not identified from observations is $I_{45}m^{-1}$;
this can vary over a limited range,
$0.5 < I_{45}m^{-1} < 2$, for virtually any mass and EOS (see
the rotating NS models of Friedman,
Ipser \& Parker 1986 and Cook, Shapiro \& Teukolsky 1992).
The stellar oblateness induced by the star's rotation
gives rise to correction terms in the nodal precession frequency also
(see  Morsink \& Stella 1999 for a post-Newtonian formula).
Their relative importance increases for high
$\nu_s$ and $\nu_\phi$.
Yet the Lense-Thirring term dominates over a
wide range of parameters,
such that a $\sim \nu_\phi^2$ dependence is expected for $\nu_{nod}$.

An approximately quadratic dependence of $\nu_{HBO}$ on the higher
frequency kHz QPOs has been measured in a number of LMXRBs.
This dependence was originally suggested on the basis of a few power
spectra of the Atoll source 4U1728-34 (Stella \& Vietri 1998a).
Ford \& van der Klis (1998) analysed a large set of power spectra from
the same source and determined
that the frequency $\nu_{low}$ of the $\sim 10-50$~Hz QPOs scales as
$\nu_2^{2.11\pm0.06}$. Note that the low frequency QPO vs. $\nu_2$
relation
of this source appears to be double-branched, with the centroid
frequency
shifting  by a factor of $\sim 2$ across different observations. This
suggests that on occasions the 2nd harmonic of $\nu_{HBO}$ is excited
instead
of the fundamental. Stella \& Vietri (1998b) first
noticed that the HBO frequency of the Z-source GX~17+2 displays a
nearly quadratic dependence on $\nu_2$.
Psaltis et al. (1999) carried out a systematic study of Z-sources
and determined that the HBO frequency
is consistent with a $\nu_2^2$ scaling (Cyg~X-2 and Sco~X-1
show evidence for a somewhat flatter dependence).
In essence these results confirmed one of the basic features of the
RPM, namely the nearly quadratic dependence
of the nodal precession frequency on the $\phi$-frequency.

If the NS spin frequency is measured, then for any value
of $\nu_\phi$ the model yields a predicted nodal precession frequency
which is uncertain only by a factor of a few, mainly due 
to the allowed range of $I_{45}m^{-1}$  (Stella \& Vietri 1998a).
Only in the Atoll source 4U1728-34
burst oscillations and simultaneous kHz QPOs and HBOs have so far been
detected unambiguously (Strohmayer et al. 1996; Ford \& van der Klis
1998;
Mendez \& van der Klis 1999).
Therefore its QPO frequencies can be used to
test both the $\nu_{HBO}$ and $\Delta\nu$ versus
$\nu_2$ relationships predicted by the RPM, when the NS spin
derived from burst oscillations is used ($\nu_{burst} \simeq 364$~Hz).
In order to take fully into account of all the effects that contribute
determining geodetic motion in the vicinity of the
NS, we adopted a numerical approach and computed
the spacetime metric of the star using Stergioulas' (1995) code, an
equivalent
of that of Cook et al.(1992); see also Stergioulas and Friedman (1995).
From this, $\nu_{r}$ and $\nu_{nod}$ were derived
as a function of $\nu_\phi$ for infinitesimally small
tilt angles and eccentricities
(Morsink \& Stella 1999; Stella, Vietri \& Morsink 1999).

Fig.~1C shows the measured values of $\Delta\nu$ and $\nu_{HBO}$
versus $\nu_2$ in 4U1728-34.
Relatively high NS masses (see also Sect.~2) and
stiff EOSs such as AU and UU (Wiringa et al. 1988)
are required in this application of the RPM.
The solid lines in Fig.~1C are for a $1.93$~M$_\odot$ NS
with EOS AU and $\nu_s=364$~Hz.
A good agreement is obtained if the HBO frequency, the
lower of the two branches seen in 4U1728-34,
is identified with the 2nd harmonics of $\nu_{nod}$ ({\it i.e.}
$2\nu_{nod}$; 
see also Morsink \& Stella 1999; Stella \& Vietri 1999a).
Correspondingly the upper HBO branch is well fit by $4\nu_{nod}$.
The geometry of tilted orbits in the innermost disk regions
might be such that a stronger signal is produced at
the even harmonics of the nodal precession
frequency (e.g. Psaltis et al. 1999).
The frequency range and trend of the epicyclic frequency $\nu_r$
in this model are also in reasonable agreement with the $\Delta\nu$
measurements; a more complex model is clearly required in order to fit
these data more accurately. 

In summary, the model presented here is capable of reproducing the
salient features of both the $\Delta\nu$ versus $\nu_2$ and $\nu_{HBO}$
versus $\nu_2$ relationships, with just two free parameters ($M$ and
the EOS), the allowed range of which is fairly limited (moreover the EOS
cannot even be varied continuously !).
Concerning Z-sources and all other Atoll sources for which burst
oscillations have not been detected yet, the NS spin can
still be regarded as a free parameter. The application of the RPM
to the HBOs of these
sources can therefore be used to constrain the spin of
their NSs. Conversely, in those Atoll source in which $\nu_{burst}$
is measured, but HBOs have not been detected yet, the RPM can be used to
predict $\nu_{HBO}$.
These issues are briefly addressed in the next Section.

\section{THE RELATIVISTIC PRECESSION MODEL AND THE PBV CORRELATION}

Psaltis, Belloni \& van der Klis (1999) recently identified two
QPOs and peaked noise components
the frequency of which follows a tight correlation over nearly three
decades.
This correlation (hereafter PBV correlation) involves both NS and BHC
LMXRBs
spanning different classes and a wide range of luminosities
(see the points in Fig.~2).
In kHz QPO NS systems, these components are the lower frequency
kHz QPOs, $\nu_1$, and the low frequency, HBO or HBO-like QPOs,
$\nu_{HBO}$.
For BHC systems and lower luminosity NS LMXRBs the correlation
involves either two QPOs, or a QPO and a peaked noise component. In all
cases
the frequency separation is about a decade and an approximate
linear relationship ($\nu_{HBO} \sim \nu_1^{0.95}$) holds.
The QPO frequencies from the peculiar NS system Cir~X-1 varies
over nearly a  decade while closely following the PBV correlation
and bridging its low and high frequency ends. Psaltis, Belloni \& van
der Klis
(1999) noted also that the $\nu_2$ vs. $\nu_1$ relations of different
Atoll
and Z-sources line-up with good accuracy.

The RPM matches precisely
the PBV correlation, without resorting to any additional assumption
(see Stella, Vietri \& Morsink 1999).
We assume that in all QPO sources, including BHCs,
$\nu_{HBO} \simeq 2\nu_{nod}$ as in 4U1728-34 (see Sect.3).
Fig.~2A shows $2\nu_{nod}$ and $\nu_\phi$ obtained from Eqs.~1-3
as a function of $\nu_{per}$ for
corotating orbits and selected values of $M$ and $a/M$.
The high frequency end of each line is dictated by
the orbital radius reaching the marginally stable orbit.

The separation of the lines in Fig.~2A testifies that while $\nu_{nod}$
depends weakly on the  mass and more strongly on $a/M$, the opposite is
true for $\nu_\phi$. By taking the
weak field ($M/r \ll 1$) and  slow rotation ($a/M \ll 1$) limit of 
Eqs.~1-3 the relevant first order dependence is made explicit,
\begin{equation}
\nu_\phi \simeq (2\pi)^{-2/5}3^{-3/5} M^{-2/5} \nu_{per}^{3/5}
\simeq 33\ m^{-2/5}\nu_{per}^{3/5} \ \ {\rm Hz} \ \ ,
\end{equation}
\begin{equation}
\nu_{nod}\simeq(2/3)^{6/5} \pi^{1/5} (a/M) M^{1/5} \nu_{per}^{6/5}
\simeq 6.7 \times 10^{-2}\ (a/M) m^{1/5} \nu_{per}^{6/5} \ \ {\rm Hz} \
\ .
\end{equation}

For the case of rotating NSs
we adopt the numerical approach outlined in Sect.~3.
Results are shown in Fig.~2B for a NS mass of
1.95~M$_{\odot}$, EOS AU
and $\nu_{s} = $~300, 600, 900 and 1200~Hz (corresponding to
$a/M =$~0.11, 0.22, 0.34 and 0.47, respectively).
Note that the approximate scalings in Eqs.~6-7
remain valid over a wide range of frequencies. Only for
the largest values of $\nu_{per}$ and $\nu_{s}$,
$\nu_{nod}$ departs substantially from the $\sim \nu_{per}^{6/5}$
dependence.

The measured QPO and peaked noise frequencies giving rise to the PBV
correlation are also plotted in Fig.~2B.
Higher kHz QPO frequencies from NS systems ($\nu_2$) are included
(for the sake of clarity NBOs and FBOs were excluded).
The agreement over the range of frequencies
spanned by each kHz QPO NS system
should not be surprising: together with the accurate matching of the
corresponding $\nu_1-\nu_2$
relationship in Z-sources, this is indeed part of the evidence
on which the RPM model was proposed.
However the fact that the dependence of $\nu_{nod}$  on $\nu_{per}$
matches
the observed $\nu_{HBO} - \nu_{1}$ correlation to a good accuracy
over $\sim 3$ decades in frequency (down to
$\nu_1$ of a few Hz), encompassing both NS and BHC systems,
provides additional independent evidence in favor of the RPM.
The observed variation of $\nu_{HBO}$ and $\nu_1$ in individual sources
(Cir~X-1 is the most striking example, see Fig.~2B)
further supports the scaling predicted by the RPM.
The matching of the observed $\nu_{2}$ vs. $\nu_{1}$ relation in terms
of $\nu_{\phi}$ vs. $\nu_{per}$ is also quite accurate.

For EOS AU and $m=1.95$, the $\nu_{HBO}$ vs. $\nu_{1}$ values of most NS
LMXRBs are best matched for $\nu_{s}$ in the $\sim 600$ to 900~Hz range.
It is apparent from Fig.~2B that $\nu_{s}$ as low as $\sim 300$~Hz are
required for the Atoll sources with $\nu_{HBO}$ somewhat below the
main PBV correlation: these are 4U1728-34 (see Sect.~3) and 4U1608-52
(from which burst oscillations have not been detected yet).
The values above are close to the
range of $\nu_s$ inferred from $\nu_{burst}$ in a number
of other Atoll sources (van der Klis 1999).
Z-type LMXRBs appear to require $\nu_{s}$ in the $\sim 600$ to 900~Hz
range, a possibility that is still open since
for none of these sources there exists yet a
$\nu_{s}$ measurement.
Note that the upper HBO branch of 4U1728-34 matches
well the main PBV correlation.
In the interpretation of Sect.~3 the lower branch corresponds to
$2\nu_{nod}$ and the upper branch to $4\nu_{nod}$.
One could further speculate that sources following
the main PBV correlation, Z-sources in particular,
are also in the upper HBO branch at $4\nu_{nod}$; in this case
their $\nu_s$ might be expected in the $\sim 300-400$~Hz
range.

\begin{figure}
\centerline{\psfig{file=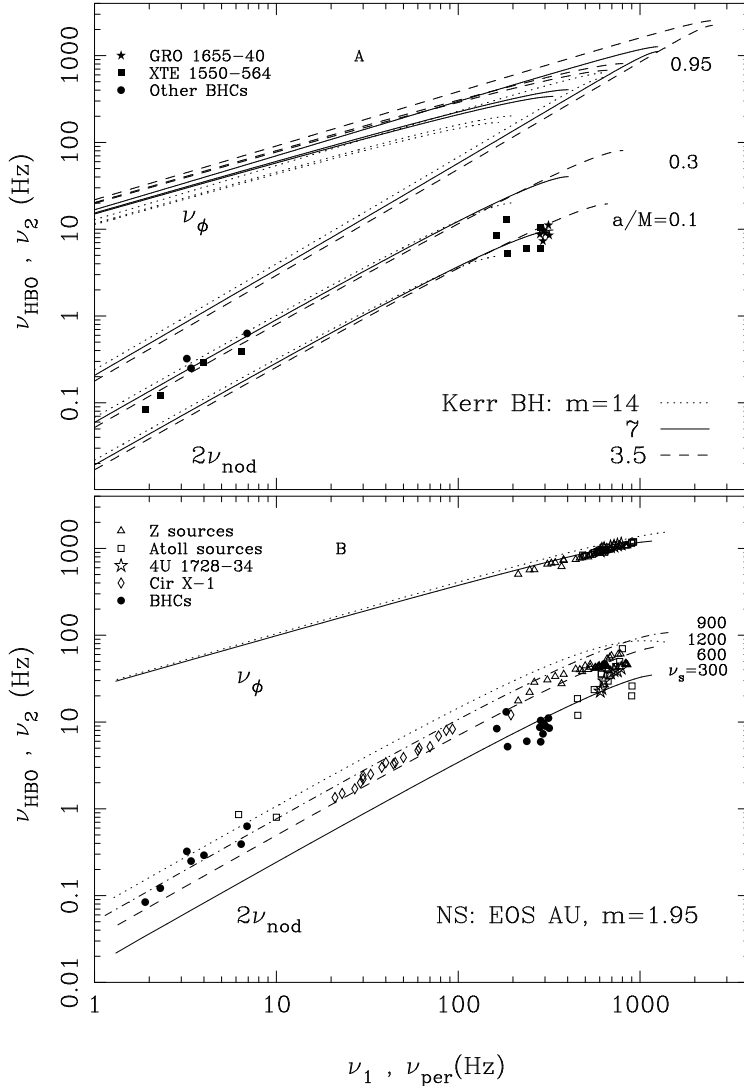, width=14cm, angle=-90}}
\caption[]{Twice the nodal precession frequency, $2\nu_{nod}$,
and $\phi$-frequency, $\nu_{\phi}$,
vs. periastron precession frequency, $\nu_{per}$, for
black hole candidates of various masses and angular momenta (panel A)
and rotating neutron star models (EOS AU, $m=1.95$) with selected
spin frequencies (panel B).
The measured QPO (or peaked noise) frequencies $\nu_1$, $\nu_2$ and
$\nu_{HBO}$
giving rise to the PBV correlation are also shown in panel B for both
BHC and NS LMXRBs and in panel A for BHC LMXRBs only; errors bars are
not plotted (see Psaltis, Belloni \& van der Klis 1999 for a complete 
list of references).
We included only those cases in which QPOs at $\nu_1$ were unambiguously
detected. NBO and FBO frequencies  are not plotted. }
\end{figure}

For BHC LMXRBs the scatter around the PBV correlation
implies values of $a/M$ of $\sim 0.1-0.3$ (see Fig.~2B).
The points from XTE~J1550-564, while inconsistent with any single value
of $a/M$, might lie along two distinct branches separated by a factor
of $\sim 2$ in $\nu_{HBO}$, similar to the case of 4U1728-34.
In the RPM the high frequency BHC QPOs
({\it e.g.} the $\sim 300$~Hz QPOs of GRO1655-40)
are interpreted terms of $\nu_1 = \nu_{per}$.
This is at variance with
the $\nu_1 = \nu_{nod}$ interpretation of Cui et al. (1998), which
requires high values of $a/M$ ($\sim 0.95$ in GRO1655-40), in contrast
with BHC accretion-driven spinup scenarios (King \& Kolb 1999).
From Fig.~2A it is apparent that the $\sim 300$~Hz QPOs from GRO~1655-40
lie close to the high frequency end of the
$a/M=0.1$, $m=7$ line. Since the mass of the BHC in GRO~1655-40
determined
through optical observations is $\sim 7$~M$_{\odot}$ (Shahbaz et al.
1999),
we conclude that, according to the RPM, $\nu_{per}\simeq 300$~Hz
close to the marginally stable orbit, where by definition
$\nu_{\phi}=\nu_{per}$. This suggests that any additional QPO
signal at $\nu_2 = \nu_{\phi}$ would be very close to
or even blended with the QPO peak at $\nu_1$.
The detection of two closeby or even partially overlapping QPO peaks
close to $\sim 300$~Hz in GRO~1655-40 would therefore provide
further evidence in favor of the RPM interpretation.

\section{DISCUSSION}

\subsection{Beat Frequency Models}

In the alternative scenario provided by 
beat frequency models, BFMs, 
disk inhomogeneities at the magnetospheric boundary ($r_m$)
and the sonic point radius ($r_s$) are accreted at the beat frequency 
between the local
Keplerian frequency, $\nu_\phi$, and the  NS spin frequency,
$\nu_s$, giving rise to the HBOs ($\nu_{HBO}= \nu_{\phi}(r_m) - \nu_s$) 
and the lower frequency kHz QPOs 
($\nu_1= \nu_{\phi}(r_s) - \nu_s$),
respectively (Alpar \& Shaham 1985; Lamb et al. 1985; Miller et al.
1998).
The higher frequency kHz QPOs are attributed to the Keplerian motion
at the sonic point radius ($\nu_2 = \nu_\phi(r_s)$).
The frequency separation $\Delta\nu$
therefore, yields the NS spin frequency, $\nu_s$.
The narrow distribution of $\nu_{s}$ inferred in this way 
($\sim 250-360$~Hz) is far from the mass shedding limit of any NS
model and  considerably less extended than that of fast MSPs 
($\sim 600$~Hz).
Accordingly the NSs of LMXRBs should be equilibrium rotators with a
narrow frequency range despite their different average mass
transfer rates, magnetic field strenghts and evolutionary histories.
Moreover, if LMXRBs are the progenitors of MSPs,
then BFMs would require a different evolutionary path leading to the 
formation of radio pulsars with $\nu_s > 400$~Hz.

The fact that in some Atoll LMXRBs 
$\nu_{burst}\simeq \Delta\nu$ (or $\simeq 2\Delta\nu$) 
is readily interpreted, because in BFMs $\nu_s \equiv \Delta\nu $. 
Yet, $\Delta\nu$ does vary and is significantly different from 
$\nu_{burst}$ in several sources (see Sect.~1). This in contrast with 
the expectations of simple BFMs. 

Attempts at fitting the PBV correlation within BFMs by using the range of 
spin frequencies inferred from $\Delta\nu$  
fail to produce a power-law like behaviour over a sufficiently large 
range of frequencies (see van der Klis 1999). This is because 
both $\nu_1$ and $\nu_{HBO}$ result from the difference of a variable 
frequency ($\nu_{\phi}$ at $r_s$ and $r_m$, respectively) and 
a fixed frequency ($\nu_s$). Moreover, 
BFMs are not applicable to BHCs, since the
{\it no hair theorem} excludes the possibility that an offset magnetic
field or radiation beam can be stably anchored to the black hole, as
required to produce the beating with the disk Keplerian frequency.

\subsection{The Relativistic Precession Model}

The RPM naturally explains the frequency range and dependence 
of the kHz QPOs and HBOs in NS LMXRBs, as well as the PBV correlation,
which involves both NS and BHC systems.
The model has a minimum number of free parameters.
Predictions that can be tested through future
analyses and/or observations include:

(a) The frequency difference $\Delta\nu=\nu_r$ is expected to decrease
also for low values of $\nu_2=\nu_\phi$ (see Eq.~4 and Fig.~1).
Moreover if the highest $\nu_2=\nu_\phi$
frequencies do originate from nearly circular orbits (see Fig.~1A), 
then $\Delta\nu$ should quickly decrease
as $\nu_2$ increases further and the orbital radius approaches
the marginally stable orbit.

(b) $\nu_2=\nu_\phi$ is expected to scale as $\nu_1^{3/5} =
\nu_{per}^{3/5}$. 
Extending the $\nu_2$ vs. $\nu_1$
correlation in NS systems toward lower frequencies
and detecting the signal at $\nu_2$ in BHC systems
would provide important new tests.

In the RPM the QPO signals are produced at
$ r/M \sim 100\ m^{-2/5}\nu_{per}^{-2/5}$, which, for
individual sources, must decrease for increasing mass accretion rates
(as $\dot M$ is positively correlated with {\it e.g.} $\nu_2$).
The inferred radii range from close to the marginally stable ($r/M \sim 6$)
to $r/M \sim 30 $ over the frequency span covered by the PBV
correlation. Many NS and BHC LMXRBs display two
component  X-ray spectra consisting of a soft
thermal component, usually interpreted in terms of emission from an optically
thick accretion disk, and a harder, often power-law like component, likely due
to a hot inner disk region. The QPOs might originate at the transition radius
between the optically thick disk and the hot inner region
as a result of occultation 
or modulated emission by inhomogeneities 
(Stella, Vietri \& Morsink 1999; see also
Di Matteo \& Psaltis 1999).

Mechanisms that can induce a finite (though small) eccentricity
and tilt in the motion of matter in the innermost disk regions 
are currently under investigation.
In the case of NSs, some kind of resonance between the star spin and the
motion of matter inhomogeneities (see {\it e.g.} Vietri \& Stella 1998)
might be responsible for the close commensurability
of $\nu_{burst}$ and $\Delta\nu$ (or $2\Delta\nu$) observed in a
few sources. These issues will be addressed elsewhere.

A simple test-particle approximation 
has been adopted so far within the RPM. 
Much needed hydrodynamical calculations  
are still in their infancy; among other 
things these are hampered by uncertainties concerning the physics of the 
innermost disk regions. Yet we note that in the hydrodynamical 
approach explored by Psaltis \& Norman (2000), 
the test particle frequencies (the same as in the RPM
plus an additional frequency at $\nu_\phi+\nu_r$)
are selected by the response of an 
annulus in the disk, when this is subject to a
wide-band input noise. 

If confirmed, the RPM will provide an unprecedented opportunity
to measure GR effects in the strong field regime, such as the
periastron precession in the vicinity of the
marginally stable orbit and the radial dependence of the Lense-Thirring
nodal precession frequency. In principle, accurately measured kHz QPO
and HBO frequencies would yield crucial information
on the compact object such as its mass and angular momentum
({\it e.g.} by solving Eqs. 1-3 for $m$, $a/M$ and $r$).
Should suitable, additional observables be found, it might
become possible to obtain a self-consistency check of the RPM,
together with tests of GR 
in the strong field regime.

\begin{acknowledgements}

I am grateful to my close collaborators in this project, S. Morsink
and M. Vietri.

\end{acknowledgements}

\end{document}